\documentclass[onecolumn,prb]{revtex4-2}
\usepackage{graphicx,psfrag,color}
\def\bc{\begin{center}}
\def\ec{\end{center}}

\def\beq{\begin{equation}}
\def\eeq{\end{equation}}

\def\d{\downarrow}
\def\u{\uparrow}

\def\br{{\bf r}}

\def\bk{{\bf k}}

\def\bc{{\bf c}}

\def\bb{{\bar b}}

\begin{document}

\title{
Josephson effect in excitonic chiral double layers
}

\author{Klaus Ziegler} 
\affiliation{
Institut f\"ur Physik, Universit\"at Augsburg, D-86135 Augsburg, Germany
}
\date{\today}

\begin{abstract}
Considering s-wave pairing in an electron-hole double layer of two chiral metals, 
we study the Josephson effect generated by a domain wall. 
The latter, which is caused by a jump of the order parameter phase,
creates two independent evanescent zero-energy modes at an exceptional point. A unique solution is found by coupling the
quasiparticle currents to the supercurrents via the continuity equation. This leads to a superposition of the zero-energy
modes, where the relative phases of the coefficients are linked to the direction of the macroscopic supercurrents.
Assuming a toroidal geometry, the effective Josephson current winds around the domain walls.
Such a system can be realized as the surface of a ring-shaped topological insulator. 
\end{abstract}

\maketitle

\section{Introduction}

Pairing of electrons or electrons and holes in solids is a fundamental effect with interesting properties such as
superconductivity 
or superfluidity. 
As an important tool to probe pairing the Josephson effect~\cite{JOSEPHSON1962251} can be employed,
which is sensitive to the phase difference of the 
pairing order parameter on both sides of a Josephson junction. It generates a macroscopic current that flows perpendicular
to the Josephson junction and which is accessible to current measurement.
This effect has been studied in many systems with electronic paired states, including conventional s-wave
superconductors~\cite{PhysRevLett.10.230}, 
unconventional and topological superconductors~\cite{PhysRevB.60.6308,PhysRevB.67.184505,Kwon2004}.
It has also been observed in atomic gases, for instance, in $^3$He \cite{Pereverzev1997}, 
in $^4$He \cite{Sukhatme2001}, and in atomic Bose-Einstein 
condensates~\cite{doi:10.1126/science.1062612,PhysRevLett.95.010402}. 

In a recent paper we considered the intralayer Josephson effect in a two-dimensional electron-electron double layer with 
interlayer s-wave pairing, assuming that both layers have a Dirac-like spectrum 
consisting of two bands and a spectral node~\cite{PhysRevLett.128.157001}. Such conditions can be realized, for instance, 
on the surface of a 3D topological insulator \cite{RevModPhys.83.1057,Burkov_2015}.
Using chiral layers with opposite chirality the formation of zero-energy edge modes plays a crucial role.
It was found that the interplay of the supercurrent and these edge modes leads to an intimate
connection of the Josephson effect and topology, which does not exist in conventional superconductors. 
It is characterized by the formation of a local current parallel to the Josephson junction, 
which competes with the conventional Josephson current that tends to cross the junction perpendicularly.
As a result of this competition we obtained an anomalous (or chiral) Josephson effect, where the direction of the 
effective Josephson current is sensitive to properties of the zero-energy edge mode. 
The latter mode is a superposition of two Majorana modes, whose
coefficients determine the direction of the Josephson current and the direction of the associated supercurrent. Since
the supercurrent is macroscopic, the current direction can be controlled easily. Thus, the interaction of the edge mode and 
the supercurrent provides a tool to control the quasiparticles.

The theory of the electronic double layer is linked to that of an electron-hole double layer (EHDL)
due to a duality transformation~\cite{2020PhRvR...2c3085S},
which connects the electronic double layer physics with excitonic physics.
In particular, the duality suggests that the intralayer Josephson effect should also exist for the 
EHDL. The electron-hole interlayer Josephson effect was studied in the presence of interlayer hopping
before by Lozovik et al.~\cite{1976JETP...44..389L,LOZOVIK1997399}. 
In that case the Josephson currents are homogeneous in each layer. 
In the following the interlayer Josephson effect will be suppressed due to the absence of interlayer hopping. 
Then we expect an intralayer Josephson effect when we implement a Josephson junction inside each layer.
Our approach is motivated by the fact that double layers of chiral materials have rich properties due to the combination 
of interlayer pairing and quasiparticle edge modes. First, the electrons in the top layer and the holes in the bottom layer interact via the attractive Coulomb interaction, which leads to the formation of excitons. Then the excitons
can condense and they form a superfluid~\cite{1976JETP...44..389L}.
A problem of chiral layers is that edge modes depend on the sample geometry due to sample boundaries. To avoid 
the contribution
of these boundaries we consider the Josephson effect on a torus. Such a geometry can be realized with two metallic 
layers, separated by a dielectric sheet~\cite{Geim2013} and connected by metallic boundaries. 
We will demonstrate for this case that the Josephson effect is robust and is determined only 
by topology, which is characterized by domain walls. 

\begin{figure}
\includegraphics[width=0.6\linewidth]{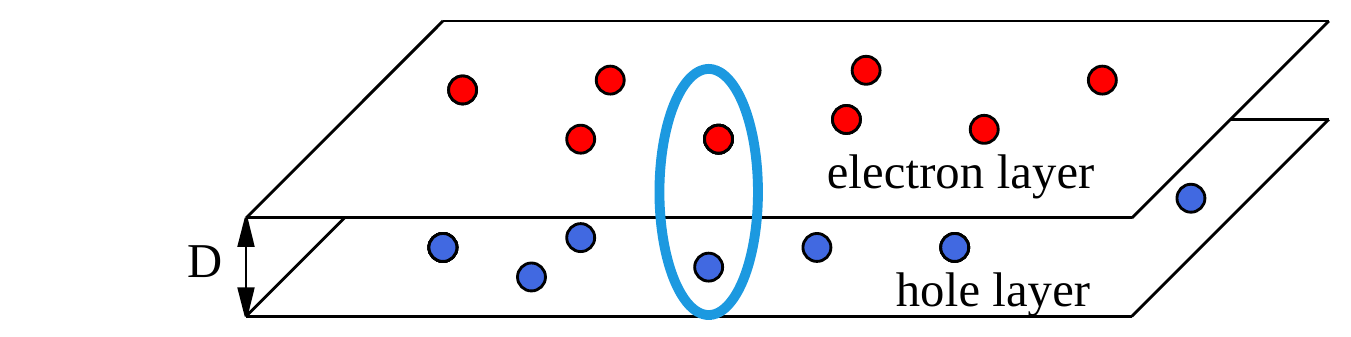}  
\caption{Electron-hole double layer, where individual pairs of electrons and holes form indirect excitons (blue ellipse)
for a sufficiently strong interlayer Coulomb interaction. The effective strength of the latter can be tuned 
by the distance $D$. These excitons can condense into a superfluid.
}
\label{fig:double_layer}
\end{figure}

\section{Model: Bogoliubov de Gennes Hamiltonian }

Our starting point is the quasiparticle Bogoliubov de Gennes (BdG) Hamiltonian that reads
\begin{equation}
H_{\rm BdG}
=\pmatrix{
H_\d & \Delta \cr
\Delta^\dagger & -H_\u^* \cr
}
,
\end{equation}
where $H_\u$ describes hopping of electrons in the top layer and $H_\d$ describes hopping of holes in the bottom layer.
$\Delta$ is the interlayer pairing order parameter, calculated in terms of the BCS mean-field 
approach~\cite{PhysRev.108.1175}. In the specific case of a chiral double layer the hopping terms $H_{\u,\d}$
carry a band index $\mu=1,2$ for two bands in each layer that is expanded in terms of Pauli matrices 
$\{\sigma_\mu\}_{\mu=1,2,3}$ with the $2\times2$ unit matrix $\sigma_0$.  We assume that $H_{\u,\d}$
are real symmetric matrices with $H_\u=H_\d=h_1\sigma_1+h_2\sigma_2$. Then we can introduce the EHDL
Hamiltonian
\begin{equation}
\label{Hamiltonian00}
H_{\rm EHDL}
=\pmatrix{
h_1\sigma_1+h_2\sigma_2 & \Delta\sigma_3 \cr
\Delta^*\sigma_3 & h_1\sigma_1+h_2\sigma_2\cr
}
\]
\[
=\pmatrix{
\sigma_0 & 0 \cr
0 & \sigma_3 \cr
}\pmatrix{
h_1\sigma_1+h_2\sigma_2 & \Delta\sigma_0 \cr
\Delta^*\sigma_0 & -h_1\sigma_1-h_2\sigma_2\cr
}
\pmatrix{
\sigma_0 & 0 \cr
0 & \sigma_3 \cr
}
.
\end{equation}
For translational invariant hopping terms $h_{1,2}$ and a uniform order parameter $\Delta$ the Fourier representation
of $H_{\rm EHDL}$ depends on the 2D wave vector $\bk=(k_x,k_y)$
and its gapped quasiparticle dispersion reads $E_\bk=\pm \sqrt{h_{1\bk}^2+h_{2\bk}^2+|\Delta|^2}$.
The latter agrees with the dispersion of the BdG Hamiltonian with layers of opposite chirality~\cite{PhysRevLett.128.157001}.
An example for a chiral system is the honeycomb lattice,
which is bipartite and consists of two triangular sublattices. In this case the sublattice index $\mu=1,2$ refers
to the two triangular lattices and the real space coordinates $\br=(x,y)$ refer only to one of the two triangular lattices.

An important feature of $H_{\rm EHDL}$ is the sign change under a particle-hole transformation
\beq
H_{\rm EHDL}\to
TH_{\rm EHDL}T
=- H_{\rm EHDL}
\ ,\ \ 
T=\pmatrix{
 \sigma_3 & 0 \cr
0 & -\sigma_3 \cr
}
,
\eeq
which reflects a chiral symmetry of quasiparticles in the paired EHDL:
\beq
U_\alpha H_{\rm EHDL} U^\dagger_\alpha
\ ,\ \ 
U_\alpha=\exp(\alpha T)
.
\eeq

\subsection{Evanescent quasiparticle modes at a domain wall}

\begin{figure}[t]
\begin{center}
\includegraphics[width=7cm,height=2cm]{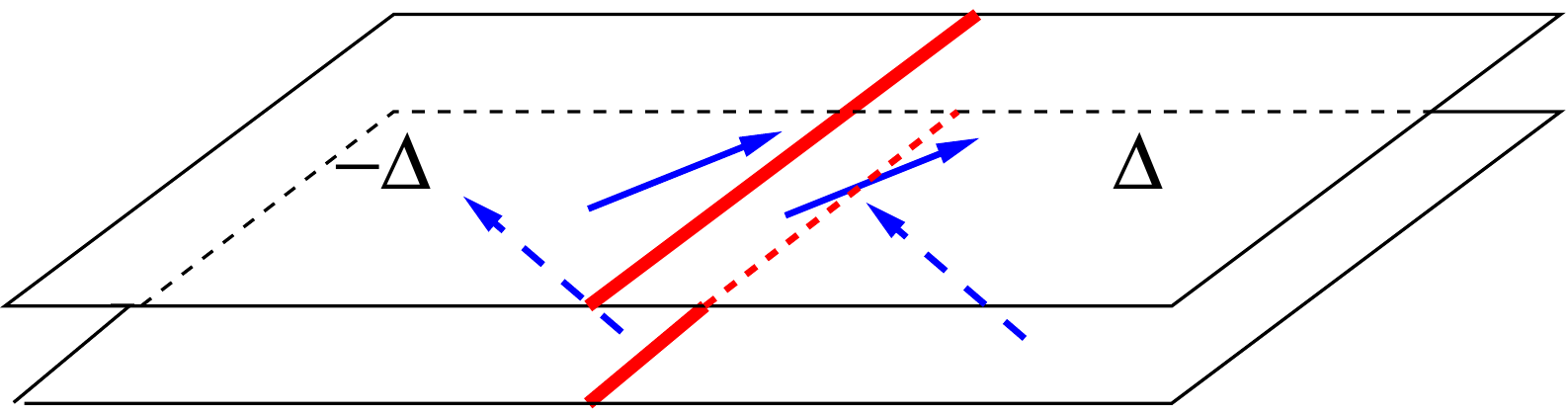}
\caption{
Electron-hole double layer with domain walls (red lines), which is given by a sign jump of the pairing order
parameter. The local quasiparticle currents (blue arrows) flow in the same (opposite) direction in the two layers 
parallel (perpendicular) to the domain walls. 
}
\label{fig:domain_wall}
\end{center}
\end{figure}

An inhomogeneous order parameter $\Delta$ breaks translational invariance
and divides the layers in different regions for different values of $\Delta$. 
Here we will consider the case in which only the phase of $\Delta$ changes in space while $|\Delta|$ is uniform.
As a Josephson junction \cite{JOSEPHSON1962251} we choose a domain wall at $x=0$ along the $y$ direction
(cf. Fig. \ref{fig:domain_wall}), where the order parameter changes.
It should be kept in mind here that the domain wall is created by a potential inside the layers, and
the corresponding change of the order parameter is obtained via the BCS-like equation.
In practice, this requires some tedious calculations~\cite{SPUNTARELLI2010111} and we simply focus here
on the domain wall in terms of $\Delta(x)$, following the recipe proposed in Ref.~[\onlinecite{2000PhRvB..6110267R}],
that it is sufficient to consider the simple case in which the phase of $\Delta$ jumps at the domain 
wall~\cite{2000PhRvB..6110267R,fradkin13}. Thus, we assume for the pairing order parameter
\beq
\label{domain_wall}
\Delta(x)=\cases{
\Delta_-=|\Delta|e^{-i\theta/2} & $x\le 0$ \cr
\Delta_+=|\Delta|e^{i\theta/2} & $x>0$ \cr
}
.
\eeq
Such a discontinuous phase change creates evanescent modes inside the gap that exists for a uniform order 
parameter~\cite{2018RSPTA.37680140S}. 

Next, we analyze the effect of the domain
wall on the low-energy approximation of the BdG Hamiltonian with $h_1\sim iv_F\hbar\partial_x$, $h_2\sim v_F\hbar k_y\sim0$
and with the Fermi velocity $v_F$.
We assume periodic boundary conditions in the $y$-direction such that an eigenmode with $k_y=0$ exists.
Then we get from Eq. (\ref{Hamiltonian00})
\begin{equation}
\label{Hamiltonian2}
H^{}_{\rm EHDL} 
=\pmatrix{
iv_F\hbar\partial_x\sigma_1+v_F\hbar k_y\sigma_2 & \Delta(x)\sigma_3 \cr
\Delta^*(x)\sigma_3 & iv_F\hbar\partial_x\sigma_1+v_F\hbar k_y\sigma_2\cr
}
.
\end{equation}
For the eigenmodes we make the ansatz $\Psi(x)=\psi e^{-bx}$,
where $\psi$ is a four-componet spinor and $b$ depends on $x$. This gives for $k_y=0$ with $\bb=v_F\hbar b$ the 
eigenmode equation
\beq
H^{}_{\rm EHDL}\Psi(x)  
=\pmatrix{
0 & -i\bb & \Delta &0\cr
-i\bb & 0 & 0 & -\Delta \cr
\Delta^* & 0 & 0 & -i\bb \cr
0 & -\Delta^* & -i\bb & 0 \cr
}
\Psi(x)
=E\Psi(x)
\label{H-matrix}
,
\eeq
which has the eigenvalue $-\sqrt{|\Delta|^2-\bb^2}$ with the eigenspinors
\beq
\label{eigenbasis_eh}
\psi_{1-}=\pmatrix{
1 \cr
0 \cr
-\sqrt{|\Delta|^2-\bb^2}/\Delta \cr
-i\bb/\Delta \cr
},\ \ 
\psi_{2-}=\pmatrix{
0 \cr
1 \cr
i\bb/\Delta \cr
\sqrt{|\Delta|^2-\bb^2}/\Delta \cr
}
\]
and the eigenvalue $\sqrt{|\Delta|^2-\bb^2}$ with the eigenspinors 
\[
\psi_{1+}=\pmatrix{
1 \cr
0 \cr
\sqrt{|\Delta|^2-\bb^2}/\Delta  \cr
-i\bb/\Delta\cr
}
,\ \
\psi_{2+}=\pmatrix{
0 \cr
1 \cr
i\bb/\Delta \cr
-\sqrt{|\Delta|^2-\bb^2}/\Delta \cr
}
.
\eeq
The $x$-dependent step function $\bb(x)$ with
\[
\bb(x)=\cases{
\bb_- & $x<0$ \cr
\bb_+ & $x\ge 0$ \cr
}
\]
must be determined by the matching conditions at the domain wall at $x=0$ as 
\beq
\label{match1}
\frac{\bb_+}{\Delta_+}=\frac{\bb_-}{\Delta_-}
,
\eeq
where the energies of these modes must be the same: $\sqrt{|\Delta|^2-\bb_-^2}=\sqrt{|\Delta|^2-\bb_+^2}$.
The latter is the case for $\bb_+^2=\bb_-^2$.
Together with Eq. (\ref{domain_wall}) this implies $\bb_-=\bb_+e^{-i\theta}$ and $\bb_+^2=|\Delta|^2$.
This is satisfied for $\theta=\pi$, which implies $\bb_-=-\bb_+$, $\Delta_\pm=\pm i|\Delta|$ and 
\beq
\frac{\bb_\pm}{\Delta_\pm}= -i\frac{\bb_+}{|\Delta|}
.
\eeq
Then we get an exponential decay on both sides of the domain wall for $\bb_\pm=\pm|\Delta|$.

It should be noted that the mean-field equation is $U(1)$ invariant, i.e., the 
global phase of the order parameter is not fixed. For instance, a global $U(1)$ transformation
$\pm i|\Delta|\to\pm|\Delta|$ gives a real pairing order parameter. 

The matching conditions lead to two zero-energy modes 
\beq
\Psi_1=
\frac{1}{{\cal N}}\pmatrix{
1 \cr
0 \cr
0 \cr
1 \cr
}e^{-|\Delta||x|/\hbar v_F}
\ ,\ \ 
\Psi_2=\frac{1}{{\cal N}}\pmatrix{
0 \cr
1 \cr
-1 \cr
0 \cr
}e^{-|\Delta||x|/\hbar v_F}
\label{zero_modes02}
\eeq
with the normalization ${\cal N}=\sqrt{2v_F\hbar/|\Delta|}$. This means that $E=0$ represents an exceptional 
point~\cite{Kato:101545}, where 
the four-dimensional eigenspace defined in Eq. (\ref{eigenbasis_eh}) coalesces to a two-dimensional eigenspace of 
zero-energy modes.
In the context of line defects in the BdG Hamiltonian the appearance of 
exceptional points has been discussed recently in Ref.~\cite{Mandal_2015}.
It should be noted that the zero-energy eigenmodes in Eq. (\ref{zero_modes02}) are real (i.e., they are Majoranas).
Therefore, any superposition of the two zero-energy modes $\Phi=a_1\Psi_1+a_2\Psi_2$ with complex coefficients
$a_j=|a_j|e^{i\varphi_j}$ (and normalization $|a_1|^2+|a_2|^2=1$) is also a zero-energy mode.
Consequently, these eigenmodes are complex in general and only real for a special choice of the coefficients.
Since both zero-energy eigenmodes in Eq. (\ref{zero_modes02}) decay exponentially with $|x|$, 
we must apply an additional condition to obtain a unique solution for the physical system. 
Based on the idea that the zero-energy eigenmodes represent a current density near the domain wall, which couples
through the Josephson effect to the macroscopic supercurrent~\cite{PhysRevLett.128.157001}, we determine a unique
solution by fixing the current.  

\section{Current conservation: continuity equations}
\label{sect:currents}

Due to the absence of interlayer tunneling, the current is conserved in each layer separately. 
We project the zero-energy mode $\Phi$ to the modes $\Phi_{\u,\d}$ of the individual layers as
\beq
\Phi=\pmatrix{
\Phi_\u \cr
\Phi_\d \cr
}
\ \ {\rm with}\ \ 
\Phi_\u=\pmatrix{
a_1 \cr
a_2 \cr
}\frac{e^{-|\Delta||x|/\hbar v_F}}{{\cal N}}
\ ,\ \ 
\Phi_\d=\pmatrix{
-a_2 \cr
a_1 \cr
}
\frac{e^{-|\Delta||x|/\hbar v_F}}{{\cal N}}
.
\eeq
Then the BdG equation~\cite{sigal22} yields the continuity equations with the quasiparticle current operator
$j_{x}=\frac{i}{\hbar}[H_{\rm EHDL},x]$. We obtain for the top layer
\beq
\label{cont_up}
\partial_t\Phi_\u\cdot\Phi_\u+\partial_xj_{x\u}
=i\frac{\Delta}{\hbar}\Psi^*_\d\sigma_3\Psi_\u-i\frac{\Delta^*}{\hbar}\Psi^*_\u\sigma_3\Psi_\d
=-\frac{i}{v_F\hbar^2}(\Delta-\Delta^*)|\Delta|Re(a_1a_2^*)e^{-2|\Delta||x|/\hbar v_F}
\eeq
and for the bottom layer
\beq
\label{cont_down}
\partial_t\Phi_\d\cdot\Phi_\d+\partial_xj_{x\d}
=i\frac{\Delta^*}{\hbar}\Psi^*_\u\sigma_3\Psi_\d-i\frac{\Delta}{\hbar}\Psi^*_\d\sigma_3\Psi_\u
=\frac{i}{v_F\hbar^2}(\Delta-\Delta^*)|\Delta|Re(a_1a_2^*)e^{-2|\Delta||x|/\hbar v_F}
,
\eeq
where $i(\Delta-\Delta^*)=-2 sgn(x)|\Delta|$. 
The expressions on the right-hand sides of the continuity equations define 
supercurrents $j^s_{x\u,\d}$ through the relation
\beq
\label{supercurrent0}
\partial_{x}j^s_{x\u,\d}
=\mp 2sgn(x)\frac{|\Delta|^2}{v_F\hbar^2}Re(a_1a_2^*)e^{-2|\Delta||x|/\hbar v_F}
.
\eeq
After the $x$--integration from the domain wall to some position $x$ this expression provides
the $x$-components of the supercurrents as
\beq
j^s_{x\u,\d}(x)=\mp\frac{ |\Delta|}{\hbar}Re(a_1a_2^*)(1-e^{-2|\Delta||x|/\hbar v_F})
,
\eeq
which vanishes at the domain wall $x=0$.
Moreover, the $x$-components of the quasiparticle currents $j_{x\u,\d}$ read
\beq
\label{qp_current_x}
j_{x\u}=-j_{x\d}
=-v_F\Phi_\u\cdot\sigma_1\Phi_\u
=-\frac{|\Delta|}{\hbar}Re(a_1a_2^*)e^{-2|\Delta||x|/\hbar v_F}
.
\eeq
The stationary case $\partial_t\Phi_\u\cdot\Phi_\u=\partial_t\Phi_\d\cdot\Phi_\d=0$ provides the condition
\[
\partial_{x}j^s_{x\u,\d}+\partial_{x}j_{x\u,\d}
=\pm 2sgn(x)\left[-\frac{|\Delta|^2}{v_F\hbar^2}Re(a_1a_2^*)e^{-2|\Delta||x|/\hbar v_F}
+\frac{|\Delta|^2}{v_F\hbar^2}Re(a_1a_2^*)e^{-2|\Delta||x|/\hbar v_F}\right] =0
\]
on both side of the domain wall. Finally, the $y$-components of the quasiparticle currents
\beq
\label{qp_current_y}
j_{y\u}=j_{y\d}=-v_F\Phi_\d\cdot\sigma_2\Phi_\d=-\frac{|\Delta|}{\hbar}Im(a_1a_2^*)e^{-2|\Delta||x|/\hbar v_F}
\eeq
yield the vorticities of the quasiparticle currents
\beq
(\nabla\times {\bf j}_{\u,\d})_z=\partial_xj_{y\u,\d}=-2sgn(x)\frac{|\Delta|^2}{v_F\hbar^2}
Im(a_1a_2^*)e^{-2|\Delta||x|/\hbar v_F}
.
\eeq
Thus, the vorticity is the same in both layers but has the opposite sign for the two sides of the domain wall.
When the modes in the top and in the bottom layer are orthogonal ($\Phi_\u\cdot\Phi_\d\propto
iIm(a_1a_2^*)|\Delta|=0$),
the quasiparticle current has no $y$-component and the vorticity vanishes. This is the case for the Majorana modes.

\begin{figure}[t]
\begin{center}
\includegraphics[width=5cm,height=3cm]{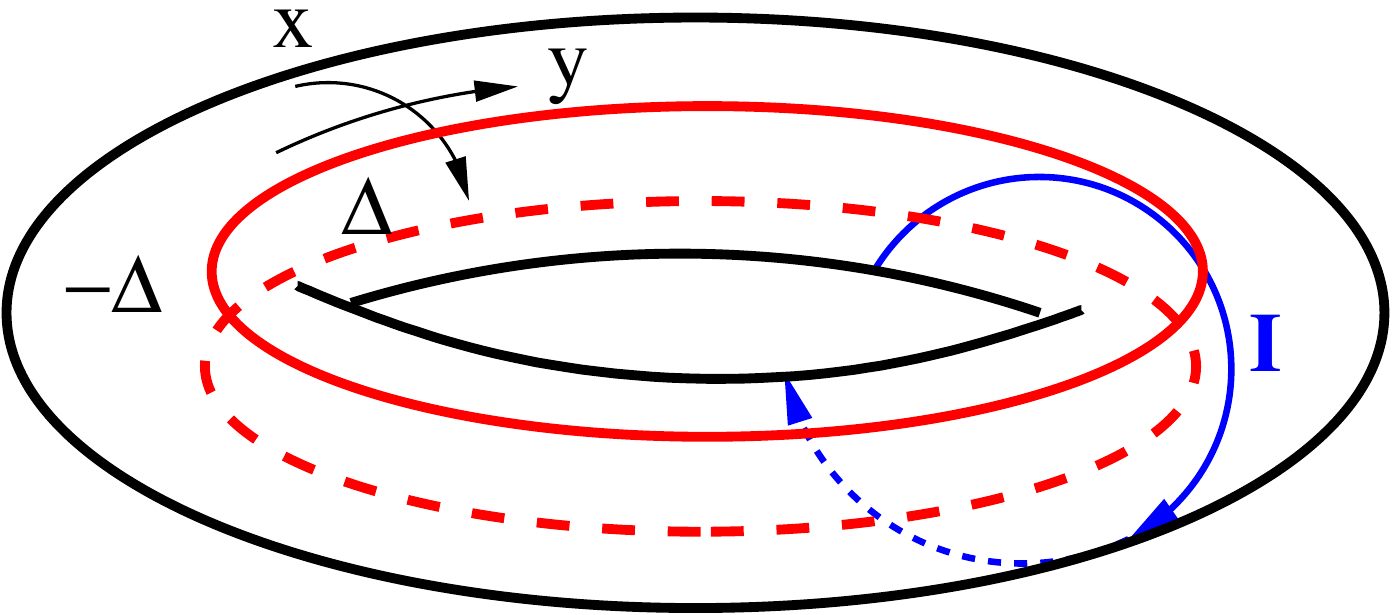}
\caption{
After gluing the double layer in Fig. \ref{fig:domain_wall} we form a torus.
Then the global current ${\bf I}$ winds along the two domain walls around the torus.
}
\label{fig:torus}
\end{center}
\end{figure}

These results reflect that the domain wall creates quasiparticle zero-energy modes, which are associated with 
local quasiparticle currents and, through the Josephson effect, are connected to the macroscopic supercurrents. 
To avoid any additional contribution of zero-energy modes we must prevent other edges, such as the geometric boundaries
in Fig. \ref{fig:domain_wall}. This is possible by making the system compact by gluing the double layers at the boundaries.
The result of this procedure is presented as a torus with two domain walls in Fig. \ref{fig:torus}. 
Such a geometry can be realized as the surface of a ring-shaped topological insulator.

\section{Discussion and Conclusion}

The results of Sect. \ref{sect:currents} reflect an intimate relationship between the macroscopic supercurrent and 
the zero-energy eigenmode of the quasiparticles via the Josephson effect. This is characterized by
the angle $\alpha$ ($-\alpha$) that exists between the current and the domain wall in the top (bottom) layer. 
Using the expressions in Eqs. (\ref{qp_current_x}) and (\ref{qp_current_y}) we obtain for the phases $\varphi_1$,
$\varphi_2$ of the eigenmode coefficients $a_1$ and $a_2$ the relation
\beq
\alpha=\arctan (j_{y\u}/j_{x\u})=-\arctan (j_{y\d}/j_{x\d})
=\frac{\pi}{2}+\varphi_2-\varphi_1
,
\eeq
such that the relative phase of the eigenmode coefficients are determined by the current direction.
This offers a direct access to this mode, provided that an external field couples to the dipole moment of the excitons
to control the supercurrent. In particular, a real (Majorana) mode appears for $\alpha=\pi/2$ when the current flows
perpendicular to the domain wall.

Our calculation was performed for the special example with a jump of the order parameter phase.
Similar calculations can be performed for other shapes of $\Delta(x)=|\Delta|e^{i\alpha(x)}$, provided that the order
parameter phase represents a kink with a global phase change $\pi$ from one boundary to the other
in $x$ direction.
Although the chiral Josephson effect will not be changed qualitatively, the decay of the
wavefunctions and of the current densities might be different. In particular, a  
broader Josephson junction might have different modes (cf. discussion in Ref. [\onlinecite{2018RSPTA.37680140S}]).

%


\begin{thebibliography}{24}%
\makeatletter
\providecommand \@ifxundefined [1]{%
 \@ifx{#1\undefined}
}%
\providecommand \@ifnum [1]{%
 \ifnum #1\expandafter \@firstoftwo
 \else \expandafter \@secondoftwo
 \fi
}%
\providecommand \@ifx [1]{%
 \ifx #1\expandafter \@firstoftwo
 \else \expandafter \@secondoftwo
 \fi
}%
\providecommand \natexlab [1]{#1}%
\providecommand \enquote  [1]{``#1''}%
\providecommand \bibnamefont  [1]{#1}%
\providecommand \bibfnamefont [1]{#1}%
\providecommand \citenamefont [1]{#1}%
\providecommand \href@noop [0]{\@secondoftwo}%
\providecommand \href [0]{\begingroup \@sanitize@url \@href}%
\providecommand \@href[1]{\@@startlink{#1}\@@href}%
\providecommand \@@href[1]{\endgroup#1\@@endlink}%
\providecommand \@sanitize@url [0]{\catcode `\\12\catcode `\$12\catcode
  `\&12\catcode `\#12\catcode `\^12\catcode `\_12\catcode `\%12\relax}%
\providecommand \@@startlink[1]{}%
\providecommand \@@endlink[0]{}%
\providecommand \url  [0]{\begingroup\@sanitize@url \@url }%
\providecommand \@url [1]{\endgroup\@href {#1}{\urlprefix }}%
\providecommand \urlprefix  [0]{URL }%
\providecommand \Eprint [0]{\href }%
\providecommand \doibase [0]{https://doi.org/}%
\providecommand \selectlanguage [0]{\@gobble}%
\providecommand \bibinfo  [0]{\@secondoftwo}%
\providecommand \bibfield  [0]{\@secondoftwo}%
\providecommand \translation [1]{[#1]}%
\providecommand \BibitemOpen [0]{}%
\providecommand \bibitemStop [0]{}%
\providecommand \bibitemNoStop [0]{.\EOS\space}%
\providecommand \EOS [0]{\spacefactor3000\relax}%
\providecommand \BibitemShut  [1]{\csname bibitem#1\endcsname}%
\let\auto@bib@innerbib\@empty
\bibitem [{\citenamefont {Josephson}(1962)}]{JOSEPHSON1962251}%
  \BibitemOpen
  \bibfield  {author} {\bibinfo {author} {\bibfnamefont {B.}~\bibnamefont
  {Josephson}},\ }\bibfield  {title} {\bibinfo {title} {Possible new effects in
  superconductive tunnelling},\ }\href
  {https://doi.org/https://doi.org/10.1016/0031-9163(62)91369-0} {\bibfield
  {journal} {\bibinfo  {journal} {Physics Letters}\ }\textbf {\bibinfo {volume}
  {1}},\ \bibinfo {pages} {251 } (\bibinfo {year} {1962})}\BibitemShut
  {NoStop}%
\bibitem [{\citenamefont {Anderson}\ and\ \citenamefont
  {Rowell}(1963)}]{PhysRevLett.10.230}%
  \BibitemOpen
  \bibfield  {author} {\bibinfo {author} {\bibfnamefont {P.~W.}\ \bibnamefont
  {Anderson}}\ and\ \bibinfo {author} {\bibfnamefont {J.~M.}\ \bibnamefont
  {Rowell}},\ }\bibfield  {title} {\bibinfo {title} {Probable observation of
  the josephson superconducting tunneling effect},\ }\href
  {https://doi.org/10.1103/PhysRevLett.10.230} {\bibfield  {journal} {\bibinfo
  {journal} {Phys. Rev. Lett.}\ }\textbf {\bibinfo {volume} {10}},\ \bibinfo
  {pages} {230} (\bibinfo {year} {1963})}\BibitemShut {NoStop}%
\bibitem [{\citenamefont {Tanaka}\ \emph {et~al.}(1999)\citenamefont {Tanaka},
  \citenamefont {Hirai}, \citenamefont {Kusakabe},\ and\ \citenamefont
  {Kashiwaya}}]{PhysRevB.60.6308}%
  \BibitemOpen
  \bibfield  {author} {\bibinfo {author} {\bibfnamefont {Y.}~\bibnamefont
  {Tanaka}}, \bibinfo {author} {\bibfnamefont {T.}~\bibnamefont {Hirai}},
  \bibinfo {author} {\bibfnamefont {K.}~\bibnamefont {Kusakabe}},\ and\
  \bibinfo {author} {\bibfnamefont {S.}~\bibnamefont {Kashiwaya}},\ }\bibfield
  {title} {\bibinfo {title} {Theory of the josephson effect in a
  superconductor/one-dimensional electron gas/superconductor junction},\ }\href
  {https://doi.org/10.1103/PhysRevB.60.6308} {\bibfield  {journal} {\bibinfo
  {journal} {Phys. Rev. B}\ }\textbf {\bibinfo {volume} {60}},\ \bibinfo
  {pages} {6308} (\bibinfo {year} {1999})}\BibitemShut {NoStop}%
\bibitem [{\citenamefont {Asano}\ \emph {et~al.}(2003)\citenamefont {Asano},
  \citenamefont {Tanaka}, \citenamefont {Sigrist},\ and\ \citenamefont
  {Kashiwaya}}]{PhysRevB.67.184505}%
  \BibitemOpen
  \bibfield  {author} {\bibinfo {author} {\bibfnamefont {Y.}~\bibnamefont
  {Asano}}, \bibinfo {author} {\bibfnamefont {Y.}~\bibnamefont {Tanaka}},
  \bibinfo {author} {\bibfnamefont {M.}~\bibnamefont {Sigrist}},\ and\ \bibinfo
  {author} {\bibfnamefont {S.}~\bibnamefont {Kashiwaya}},\ }\bibfield  {title}
  {\bibinfo {title} {Josephson current in
  s-wave-superconductor/${\mathrm{sr}}_{2}{\mathrm{ruo}}_{4}$ junctions},\
  }\href {https://doi.org/10.1103/PhysRevB.67.184505} {\bibfield  {journal}
  {\bibinfo  {journal} {Phys. Rev. B}\ }\textbf {\bibinfo {volume} {67}},\
  \bibinfo {pages} {184505} (\bibinfo {year} {2003})}\BibitemShut {NoStop}%
\bibitem [{\citenamefont {Kwon}\ \emph {et~al.}(2004)\citenamefont {Kwon},
  \citenamefont {Sengupta},\ and\ \citenamefont {Yakovenko}}]{Kwon2004}%
  \BibitemOpen
  \bibfield  {author} {\bibinfo {author} {\bibfnamefont {H.-J.}\ \bibnamefont
  {Kwon}}, \bibinfo {author} {\bibfnamefont {K.}~\bibnamefont {Sengupta}},\
  and\ \bibinfo {author} {\bibfnamefont {V.~M.}\ \bibnamefont {Yakovenko}},\
  }\bibfield  {title} {\bibinfo {title} {Fractional ac josephson effect in p-
  and d-wave superconductors},\ }\href
  {https://doi.org/10.1140/epjb/e2004-00066-4} {\bibfield  {journal} {\bibinfo
  {journal} {The European Physical Journal B - Condensed Matter and Complex
  Systems}\ }\textbf {\bibinfo {volume} {37}},\ \bibinfo {pages} {349}
  (\bibinfo {year} {2004})}\BibitemShut {NoStop}%
\bibitem [{\citenamefont {Pereverzev}\ \emph {et~al.}(1997)\citenamefont
  {Pereverzev}, \citenamefont {Loshak}, \citenamefont {Backhaus}, \citenamefont
  {Davis},\ and\ \citenamefont {Packard}}]{Pereverzev1997}%
  \BibitemOpen
  \bibfield  {author} {\bibinfo {author} {\bibfnamefont {S.~V.}\ \bibnamefont
  {Pereverzev}}, \bibinfo {author} {\bibfnamefont {A.}~\bibnamefont {Loshak}},
  \bibinfo {author} {\bibfnamefont {S.}~\bibnamefont {Backhaus}}, \bibinfo
  {author} {\bibfnamefont {J.~C.}\ \bibnamefont {Davis}},\ and\ \bibinfo
  {author} {\bibfnamefont {R.~E.}\ \bibnamefont {Packard}},\ }\bibfield
  {title} {\bibinfo {title} {Quantum oscillations between two weakly coupled
  reservoirs of superfluid 3he},\ }\href {https://doi.org/10.1038/41277}
  {\bibfield  {journal} {\bibinfo  {journal} {Nature}\ }\textbf {\bibinfo
  {volume} {388}},\ \bibinfo {pages} {449} (\bibinfo {year}
  {1997})}\BibitemShut {NoStop}%
\bibitem [{\citenamefont {Sukhatme}\ \emph {et~al.}(2001)\citenamefont
  {Sukhatme}, \citenamefont {Mukharsky}, \citenamefont {Chui},\ and\
  \citenamefont {Pearson}}]{Sukhatme2001}%
  \BibitemOpen
  \bibfield  {author} {\bibinfo {author} {\bibfnamefont {K.}~\bibnamefont
  {Sukhatme}}, \bibinfo {author} {\bibfnamefont {Y.}~\bibnamefont {Mukharsky}},
  \bibinfo {author} {\bibfnamefont {T.}~\bibnamefont {Chui}},\ and\ \bibinfo
  {author} {\bibfnamefont {D.}~\bibnamefont {Pearson}},\ }\bibfield  {title}
  {\bibinfo {title} {Observation of the ideal josephson effect in superfluid
  4he},\ }\href {https://doi.org/10.1038/35077024} {\bibfield  {journal}
  {\bibinfo  {journal} {Nature}\ }\textbf {\bibinfo {volume} {411}},\ \bibinfo
  {pages} {280} (\bibinfo {year} {2001})}\BibitemShut {NoStop}%
\bibitem [{\citenamefont {Cataliotti}\ \emph {et~al.}(2001)\citenamefont
  {Cataliotti}, \citenamefont {Burger}, \citenamefont {Fort}, \citenamefont
  {Maddaloni}, \citenamefont {Minardi}, \citenamefont {Trombettoni},
  \citenamefont {Smerzi},\ and\ \citenamefont
  {Inguscio}}]{doi:10.1126/science.1062612}%
  \BibitemOpen
  \bibfield  {author} {\bibinfo {author} {\bibfnamefont {F.~S.}\ \bibnamefont
  {Cataliotti}}, \bibinfo {author} {\bibfnamefont {S.}~\bibnamefont {Burger}},
  \bibinfo {author} {\bibfnamefont {C.}~\bibnamefont {Fort}}, \bibinfo {author}
  {\bibfnamefont {P.}~\bibnamefont {Maddaloni}}, \bibinfo {author}
  {\bibfnamefont {F.}~\bibnamefont {Minardi}}, \bibinfo {author} {\bibfnamefont
  {A.}~\bibnamefont {Trombettoni}}, \bibinfo {author} {\bibfnamefont
  {A.}~\bibnamefont {Smerzi}},\ and\ \bibinfo {author} {\bibfnamefont
  {M.}~\bibnamefont {Inguscio}},\ }\bibfield  {title} {\bibinfo {title}
  {Josephson junction arrays with bose-einstein condensates},\ }\href
  {https://doi.org/10.1126/science.1062612} {\bibfield  {journal} {\bibinfo
  {journal} {Science}\ }\textbf {\bibinfo {volume} {293}},\ \bibinfo {pages}
  {843} (\bibinfo {year} {2001})},\ \Eprint
  {https://arxiv.org/abs/https://www.science.org/doi/pdf/10.1126/science.1062612}
  {https://www.science.org/doi/pdf/10.1126/science.1062612} \BibitemShut
  {NoStop}%
\bibitem [{\citenamefont {Albiez}\ \emph {et~al.}(2005)\citenamefont {Albiez},
  \citenamefont {Gati}, \citenamefont {F\"olling}, \citenamefont {Hunsmann},
  \citenamefont {Cristiani},\ and\ \citenamefont
  {Oberthaler}}]{PhysRevLett.95.010402}%
  \BibitemOpen
  \bibfield  {author} {\bibinfo {author} {\bibfnamefont {M.}~\bibnamefont
  {Albiez}}, \bibinfo {author} {\bibfnamefont {R.}~\bibnamefont {Gati}},
  \bibinfo {author} {\bibfnamefont {J.}~\bibnamefont {F\"olling}}, \bibinfo
  {author} {\bibfnamefont {S.}~\bibnamefont {Hunsmann}}, \bibinfo {author}
  {\bibfnamefont {M.}~\bibnamefont {Cristiani}},\ and\ \bibinfo {author}
  {\bibfnamefont {M.~K.}\ \bibnamefont {Oberthaler}},\ }\bibfield  {title}
  {\bibinfo {title} {Direct observation of tunneling and nonlinear
  self-trapping in a single bosonic josephson junction},\ }\href
  {https://doi.org/10.1103/PhysRevLett.95.010402} {\bibfield  {journal}
  {\bibinfo  {journal} {Phys. Rev. Lett.}\ }\textbf {\bibinfo {volume} {95}},\
  \bibinfo {pages} {010402} (\bibinfo {year} {2005})}\BibitemShut {NoStop}%
\bibitem [{\citenamefont {Ziegler}\ \emph {et~al.}(2022)\citenamefont
  {Ziegler}, \citenamefont {Sinner},\ and\ \citenamefont
  {Lozovik}}]{PhysRevLett.128.157001}%
  \BibitemOpen
  \bibfield  {author} {\bibinfo {author} {\bibfnamefont {K.}~\bibnamefont
  {Ziegler}}, \bibinfo {author} {\bibfnamefont {A.}~\bibnamefont {Sinner}},\
  and\ \bibinfo {author} {\bibfnamefont {Y.~E.}\ \bibnamefont {Lozovik}},\
  }\bibfield  {title} {\bibinfo {title} {Anomalous josephson effect of $s$-wave
  pairing states in chiral double layers},\ }\href
  {https://doi.org/10.1103/PhysRevLett.128.157001} {\bibfield  {journal}
  {\bibinfo  {journal} {Phys. Rev. Lett.}\ }\textbf {\bibinfo {volume} {128}},\
  \bibinfo {pages} {157001} (\bibinfo {year} {2022})}\BibitemShut {NoStop}%
\bibitem [{\citenamefont {Qi}\ and\ \citenamefont
  {Zhang}(2011)}]{RevModPhys.83.1057}%
  \BibitemOpen
  \bibfield  {author} {\bibinfo {author} {\bibfnamefont {X.-L.}\ \bibnamefont
  {Qi}}\ and\ \bibinfo {author} {\bibfnamefont {S.-C.}\ \bibnamefont {Zhang}},\
  }\bibfield  {title} {\bibinfo {title} {Topological insulators and
  superconductors},\ }\href {https://doi.org/10.1103/RevModPhys.83.1057}
  {\bibfield  {journal} {\bibinfo  {journal} {Rev. Mod. Phys.}\ }\textbf
  {\bibinfo {volume} {83}},\ \bibinfo {pages} {1057} (\bibinfo {year}
  {2011})}\BibitemShut {NoStop}%
\bibitem [{\citenamefont {Burkov}(2015)}]{Burkov_2015}%
  \BibitemOpen
  \bibfield  {author} {\bibinfo {author} {\bibfnamefont {A.~A.}\ \bibnamefont
  {Burkov}},\ }\bibfield  {title} {\bibinfo {title} {Chiral anomaly and
  transport in weyl metals},\ }\href
  {https://doi.org/10.1088/0953-8984/27/11/113201} {\bibfield  {journal}
  {\bibinfo  {journal} {Journal of Physics: Condensed Matter}\ }\textbf
  {\bibinfo {volume} {27}},\ \bibinfo {pages} {113201} (\bibinfo {year}
  {2015})}\BibitemShut {NoStop}%
\bibitem [{\citenamefont {{Sinner}}\ \emph {et~al.}(2020)\citenamefont
  {{Sinner}}, \citenamefont {{Lozovik}},\ and\ \citenamefont
  {{Ziegler}}}]{2020PhRvR...2c3085S}%
  \BibitemOpen
  \bibfield  {author} {\bibinfo {author} {\bibfnamefont {A.}~\bibnamefont
  {{Sinner}}}, \bibinfo {author} {\bibfnamefont {Y.~E.}\ \bibnamefont
  {{Lozovik}}},\ and\ \bibinfo {author} {\bibfnamefont {K.}~\bibnamefont
  {{Ziegler}}},\ }\bibfield  {title} {\bibinfo {title} {{Pairing transition in
  a double layer with interlayer Coulomb repulsion}},\ }\href
  {https://doi.org/10.1103/PhysRevResearch.2.033085} {\bibfield  {journal}
  {\bibinfo  {journal} {Physical Review Research}\ }\textbf {\bibinfo {volume}
  {2}},\ \bibinfo {eid} {033085} (\bibinfo {year} {2020})},\ \Eprint
  {https://arxiv.org/abs/1912.13257} {arXiv:1912.13257 [cond-mat.supr-con]}
  \BibitemShut {NoStop}%
\bibitem [{\citenamefont {{Lozovik}}\ and\ \citenamefont
  {{Yudson}}(1976)}]{1976JETP...44..389L}%
  \BibitemOpen
  \bibfield  {author} {\bibinfo {author} {\bibfnamefont {Y.~E.}\ \bibnamefont
  {{Lozovik}}}\ and\ \bibinfo {author} {\bibfnamefont {V.~I.}\ \bibnamefont
  {{Yudson}}},\ }\bibfield  {title} {\bibinfo {title} {{A new mechanism for
  superconductivity: pairing between spatially separated electrons and
  holes}},\ }\href@noop {} {\bibfield  {journal} {\bibinfo  {journal} {Soviet
  Journal of Experimental and Theoretical Physics}\ }\textbf {\bibinfo {volume}
  {44}},\ \bibinfo {pages} {389} (\bibinfo {year} {1976})}\BibitemShut
  {NoStop}%
\bibitem [{\citenamefont {Lozovik}\ and\ \citenamefont
  {Poushnov}(1997)}]{LOZOVIK1997399}%
  \BibitemOpen
  \bibfield  {author} {\bibinfo {author} {\bibfnamefont {Y.~E.}\ \bibnamefont
  {Lozovik}}\ and\ \bibinfo {author} {\bibfnamefont {A.~V.}\ \bibnamefont
  {Poushnov}},\ }\bibfield  {title} {\bibinfo {title} {Magnetism and josephson
  effect in a coupled quantum well electron-hole system},\ }\href
  {https://doi.org/https://doi.org/10.1016/S0375-9601(97)00133-3} {\bibfield
  {journal} {\bibinfo  {journal} {Physics Letters A}\ }\textbf {\bibinfo
  {volume} {228}},\ \bibinfo {pages} {399} (\bibinfo {year}
  {1997})}\BibitemShut {NoStop}%
\bibitem [{\citenamefont {Geim}\ and\ \citenamefont
  {Grigorieva}(2013)}]{Geim2013}%
  \BibitemOpen
  \bibfield  {author} {\bibinfo {author} {\bibfnamefont {A.~K.}\ \bibnamefont
  {Geim}}\ and\ \bibinfo {author} {\bibfnamefont {I.~V.}\ \bibnamefont
  {Grigorieva}},\ }\bibfield  {title} {\bibinfo {title} {Van der waals
  heterostructures},\ }\href {https://doi.org/10.1038/nature12385} {\bibfield
  {journal} {\bibinfo  {journal} {Nature}\ }\textbf {\bibinfo {volume} {499}},\
  \bibinfo {pages} {419} (\bibinfo {year} {2013})}\BibitemShut {NoStop}%
\bibitem [{\citenamefont {Bardeen}\ \emph {et~al.}(1957)\citenamefont
  {Bardeen}, \citenamefont {Cooper},\ and\ \citenamefont
  {Schrieffer}}]{PhysRev.108.1175}%
  \BibitemOpen
  \bibfield  {author} {\bibinfo {author} {\bibfnamefont {J.}~\bibnamefont
  {Bardeen}}, \bibinfo {author} {\bibfnamefont {L.~N.}\ \bibnamefont
  {Cooper}},\ and\ \bibinfo {author} {\bibfnamefont {J.~R.}\ \bibnamefont
  {Schrieffer}},\ }\bibfield  {title} {\bibinfo {title} {Theory of
  superconductivity},\ }\href {https://doi.org/10.1103/PhysRev.108.1175}
  {\bibfield  {journal} {\bibinfo  {journal} {Phys. Rev.}\ }\textbf {\bibinfo
  {volume} {108}},\ \bibinfo {pages} {1175} (\bibinfo {year}
  {1957})}\BibitemShut {NoStop}%
\bibitem [{\citenamefont {Spuntarelli}\ \emph {et~al.}(2010)\citenamefont
  {Spuntarelli}, \citenamefont {Pieri},\ and\ \citenamefont
  {Strinati}}]{SPUNTARELLI2010111}%
  \BibitemOpen
  \bibfield  {author} {\bibinfo {author} {\bibfnamefont {A.}~\bibnamefont
  {Spuntarelli}}, \bibinfo {author} {\bibfnamefont {P.}~\bibnamefont {Pieri}},\
  and\ \bibinfo {author} {\bibfnamefont {G.}~\bibnamefont {Strinati}},\
  }\bibfield  {title} {\bibinfo {title} {Solution of the bogoliubov–de gennes
  equations at zero temperature throughout the bcs–bec crossover: Josephson
  and related effects},\ }\href
  {https://doi.org/https://doi.org/10.1016/j.physrep.2009.12.005} {\bibfield
  {journal} {\bibinfo  {journal} {Physics Reports}\ }\textbf {\bibinfo {volume}
  {488}},\ \bibinfo {pages} {111} (\bibinfo {year} {2010})}\BibitemShut
  {NoStop}%
\bibitem [{\citenamefont {{Read}}\ and\ \citenamefont
  {{Green}}(2000)}]{2000PhRvB..6110267R}%
  \BibitemOpen
  \bibfield  {author} {\bibinfo {author} {\bibfnamefont {N.}~\bibnamefont
  {{Read}}}\ and\ \bibinfo {author} {\bibfnamefont {D.}~\bibnamefont
  {{Green}}},\ }\bibfield  {title} {\bibinfo {title} {{Paired states of
  fermions in two dimensions with breaking of parity and time-reversal
  symmetries and the fractional quantum Hall effect}},\ }\href
  {https://doi.org/10.1103/PhysRevB.61.10267} {\bibfield  {journal} {\bibinfo
  {journal} {\prb}\ }\textbf {\bibinfo {volume} {61}},\ \bibinfo {pages}
  {10267} (\bibinfo {year} {2000})}\BibitemShut {NoStop}%
\bibitem [{\citenamefont {Fradkin}(2013)}]{fradkin13}%
  \BibitemOpen
  \bibfield  {author} {\bibinfo {author} {\bibfnamefont {E.}~\bibnamefont
  {Fradkin}},\ }\href {https://books.google.de/books?id=x7\_6MX4ye\_wC} {\emph
  {\bibinfo {title} {Field Theories of Condensed Matter Physics}}},\ Field
  Theories of Condensed Matter Physics\ (\bibinfo  {publisher} {Cambridge
  University Press},\ \bibinfo {year} {2013})\BibitemShut {NoStop}%
\bibitem [{\citenamefont {{Sauls}}(2018)}]{2018RSPTA.37680140S}%
  \BibitemOpen
  \bibfield  {author} {\bibinfo {author} {\bibfnamefont {J.~A.}\ \bibnamefont
  {{Sauls}}},\ }\bibfield  {title} {\bibinfo {title} {{Andreev bound states and
  their signatures}},\ }\href {https://doi.org/10.1098/rsta.2018.0140}
  {\bibfield  {journal} {\bibinfo  {journal} {Philosophical Transactions of the
  Royal Society of London Series A}\ }\textbf {\bibinfo {volume} {376}},\
  \bibinfo {eid} {20180140} (\bibinfo {year} {2018})},\ \Eprint
  {https://arxiv.org/abs/1805.11069} {1805.11069 [cond-mat.supr-con]}
  \BibitemShut {NoStop}%
\bibitem [{\citenamefont {Kato}(1976)}]{Kato:101545}%
  \BibitemOpen
  \bibfield  {author} {\bibinfo {author} {\bibfnamefont {T.}~\bibnamefont
  {Kato}},\ }\href {https://cds.cern.ch/record/101545} {\emph {\bibinfo {title}
  {{Perturbation theory for linear operators; 2nd ed.}}}},\ Grundlehren der
  mathematischen Wissenschaften A Series of Comprehensive Studies in
  Mathematics\ (\bibinfo  {publisher} {Springer},\ \bibinfo {address}
  {Berlin},\ \bibinfo {year} {1976})\BibitemShut {NoStop}%
\bibitem [{\citenamefont {Mandal}(2015)}]{Mandal_2015}%
  \BibitemOpen
  \bibfield  {author} {\bibinfo {author} {\bibfnamefont {I.}~\bibnamefont
  {Mandal}},\ }\bibfield  {title} {\bibinfo {title} {Exceptional points for
  chiral majorana fermions in arbitrary dimensions},\ }\href
  {https://doi.org/10.1209/0295-5075/110/67005} {\bibfield  {journal} {\bibinfo
   {journal} {{EPL} (Europhysics Letters)}\ }\textbf {\bibinfo {volume}
  {110}},\ \bibinfo {pages} {67005} (\bibinfo {year} {2015})}\BibitemShut
  {NoStop}%
\bibitem [{\citenamefont {Sigal}(2022)}]{sigal22}%
  \BibitemOpen
  \bibfield  {author} {\bibinfo {author} {\bibfnamefont {I.}~\bibnamefont
  {Sigal}},\ }\bibfield  {title} {\bibinfo {title} {Differential equations of
  quantum mechanics},\ }\href
  {https://doi.org/https://doi.org/10.1090/qam/1611} {\bibfield  {journal}
  {\bibinfo  {journal} {Quart. Appl. Math.}\ }\textbf {\bibinfo {volume}
  {80}},\ \bibinfo {pages} {451} (\bibinfo {year} {2022})}\BibitemShut
  {NoStop}%
\end{thebibliography}
\end{document}